\documentclass[12pt]{article}
\usepackage{pdproc,epsfig}

\begin{document}

\renewcommand{\thefootnote}{\alph{footnote}}

\title{NEUTRINO MASS MODELS AND THE IMPLICATIONS OF A NON-ZERO REACTOR ANGLE}

\author{S.F.KING}

\address{ School of Physics and Astronomy,\\
University of Southampton,\\ Southampton SO17~1BJ, UK\\
 {\rm E-mail: king@soton.ac.uk}}

\abstract{In this talk we survey some of the recent promising
developments in the search for the theory behind neutrino mass and
mixing, and indeed all fermion masses and mixing. The talk is
organized in terms of a neutrino mass models decision tree according to
which the answers to experimental questions provide sign posts to
guide us through the maze of theoretical models eventually towards
a complete theory of flavour and unification. We also discuss the theoretical
implications of the measurement of a non-zero reactor angle, as hinted at by
recent experimental measurements.}

\normalsize\baselineskip=15pt

\section{Introduction}
It has been one of the long standing goals of theories of particle
physics beyond the Standard Model (SM) to predict quark and lepton
masses and mixings. With the discovery of neutrino mass and
mixing, this quest has received a massive impetus. Indeed, perhaps
the greatest advance in particle physics over the past decade has
been the discovery of neutrino mass and mixing involving two large
mixing angles commonly known as the atmospheric angle
$\theta_{23}$ and the solar angle $\theta_{12}$, while the
remaining mixing angle $\theta_{13}$, although unmeasured, is
constrained to be relatively small. The largeness of the two large
lepton mixing angles contrasts sharply with the smallness of the
quark mixing angles, and this observation, together with the
smallness of neutrino masses, provides new and tantalizing clues
in the search for the origin of quark and lepton flavour. However,
before trying to address such questions, it is worth recalling why
neutrino mass forces us to go beyond the SM.

\section{Why go beyond the Standard Model?}
Neutrino mass is zero in the SM for three independent reasons:
\begin{enumerate}
\item There are no right-handed neutrinos $\nu_R$. \item There are
only Higgs doublets of $SU(2)_L$. \item There are only
renormalizable terms.
\end{enumerate}
In the SM these conditions all apply and so neutrinos are massless
with $\nu_e$, $\nu_{\mu}$, $\nu_{\tau}$ distinguished by separate
lepton numbers $L_e$, $L_{\mu}$, $L_{\tau}$. Neutrinos and
antineutrinos are distinguished by total conserved lepton number
$L=L_e+L_{\mu}+L_{\tau}$. To generate neutrino mass we must relax
one or more of these conditions. For example, by adding
right-handed neutrinos the Higgs mechanism of the Standard Model
can give neutrinos the same type of mass as the electron mass or
other charged lepton and quark masses. It is clear that the {\it
status quo} of staying within the SM, as it is usually defined, is
not an option, but in what direction should we go?

\section{A decision tree}
\begin{figure}[h]
\begin{center}
\vspace*{13pt}
%\leftline{\hfill\vbox{\hrule width 5cm height0.001pt}\hfill}
      \mbox{\epsfig{figure=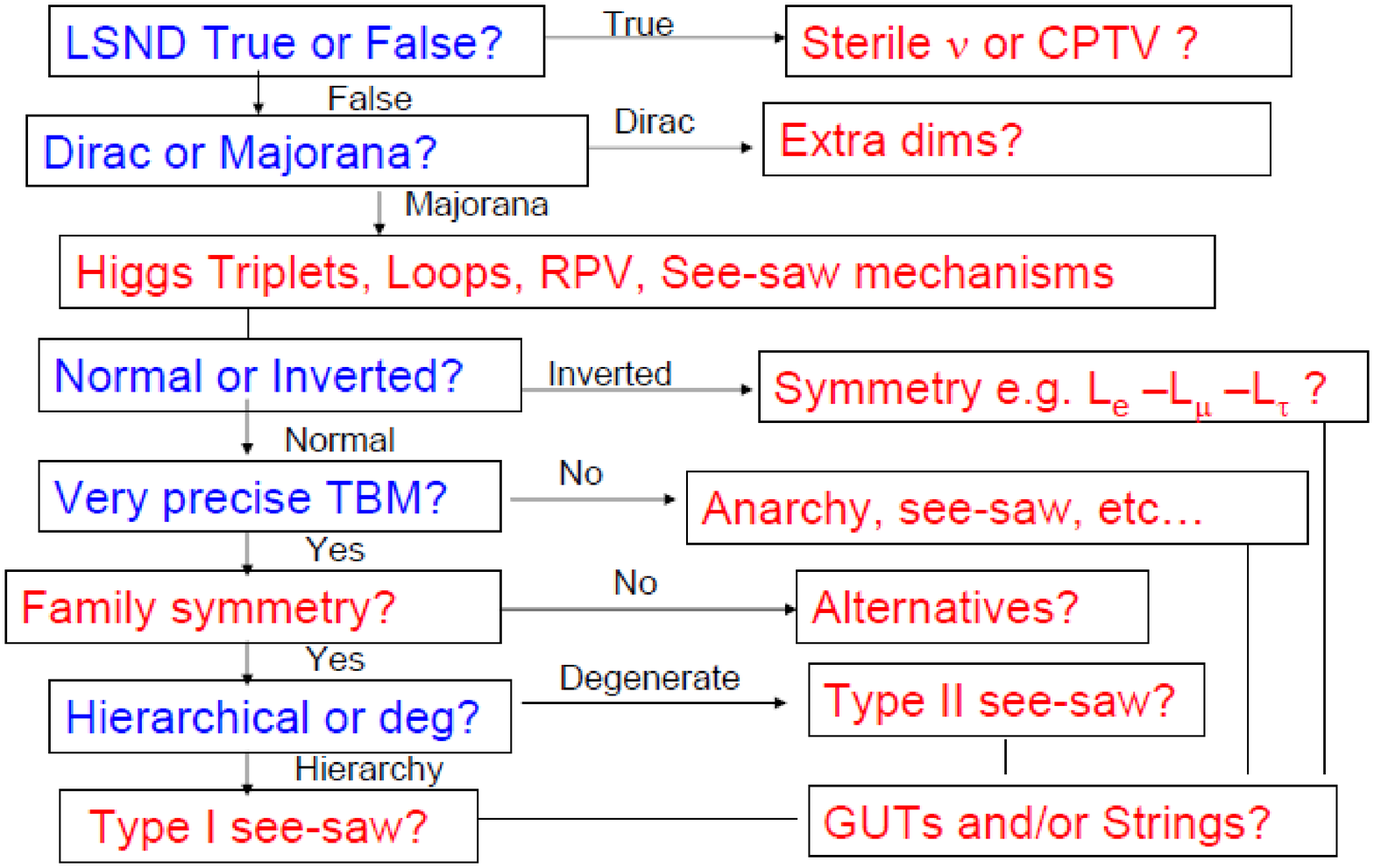,width=12.0cm}}
%\vspace*{1.4truein}     %ORIGINAL SIZE=1.6TRUEIN x 100% - 0.2TRUEIN
%\leftline{\hfill\vbox{\hrule width 5cm height0.001pt}\hfill}
\caption{\label{roadmap}Neutrino mass models decision tree.}
\end{center}
\end{figure}

This talk will be organized according to the decision tree in
Fig.\ref{roadmap}. Such a decision tree is clearly not unique
(everyone can come up with her or his personal decision tree). The
decision tree in Fig.\ref{roadmap} contains key experimental
questions (in blue) which serve as signposts along the way,
leading in particular theoretical directions, starting from the
top left hand corner with the question ``LSND True or False?''

\section{LSND True or False?}
The results from MiniBOONE do not support the LSND result, but are
consistent with the three active neutrino oscillation paradigm. If
LSND were correct then this could imply either sterile neutrinos
and/or CPT violation, or something more exotic. For the remainder
of this talk we shall assume that LSND is false, and focus on
models without sterile neutrinos.

\section{Dirac or Majorana?}
Majorana neutrino masses are of the form
$m_{LL}^{\nu}\overline{\nu_L}\nu_L^c$ where $\nu_L$ is a
left-handed neutrino field and $\nu_L^c$ is the CP conjugate of a
left-handed neutrino field, in other words a right-handed
antineutrino field. Such Majorana masses are possible since both
the neutrino and the antineutrino are electrically neutral. Such
Majorana neutrino masses violate total lepton number $L$
conservation, so the neutrino is equal to its own antiparticle. If
we introduce right-handed neutrino fields then there are two sorts
of additional neutrino mass terms that are possible. There are
additional Majorana masses of the form
$M_{RR}^{\nu}\overline{\nu_R}\nu_R^c$. In addition there are Dirac
masses of the form $m_{LR}^{\nu}\overline{\nu_L}\nu_R$. Such Dirac
mass terms conserve total lepton number $L$, but violate separate
lepton numbers $L_e, L_{\mu}, L_{\tau}$. The question of ``Dirac
or Majorana?'' is a key experimental question which could be
decided by the experiments which measure neutrino masses directly.

\section{What if Neutrinos are Dirac?}
Introducing right-handed neutrinos $\nu_R$ into the SM (with zero
Majorana mass) we can generate a Dirac neutrino mass from a
coupling to the Higgs: $\lambda_{\nu} <H>\overline{\nu_L}\nu_R
\equiv m_{LR}^{\nu}\overline{\nu_L}\nu_R$, where $<H>\approx  175$
GeV is the Higgs vacuum expectation value (VEV). A physical
neutrino mass of $m_{LR}^{\nu}\approx 0.2$ eV implies
$\lambda_{\nu}\approx 10^{-12}$. The question is why are such
neutrino Yukawa couplings so small, even compared to the charged
fermion Yukawa couplings? One possibility for small Dirac masses
comes from the idea of extra dimensions motivated by theoretical
attempts to extend the Standard Model to include gravity .

For the case of ``flat'' extra dimensions, ``compactified'' on
circles of small radius $R$ so that they are not normally
observable, it has been suggested that right-handed neutrinos (but
not the rest of the Standard Model particles) experience one or
more of these extra dimensions \cite{Antoniadis:2005aq}.
%The right handed neutrinos then only spend part of their time in our world,
%leading to very small Dirac neutrino masses
%\cite{Arkani-Hamed:1998vp}.
%In such theories there is a relation between the usual four dimensional Planck mass
%$M_{Planck}\sim 10^{19}\ GeV/c^2$, the string scale $M_{string}$ and the compactification
%radius of the ``flat'' extra dimensions $R$ given by:
%\begin{equation}
%M_{Planck}^2=M_{string}^{2+n}R^n
%\label{flat}
%\end{equation}
%where there are $n$ extra dimensions.
For example, for one extra dimension the right-handed neutrino
wavefunction spreads out over the extra dimension $R$, leading to
a suppressed Higgs interaction with the left-handed neutrino.
%with a suppression factor of $1/\sqrt{M_{string}R}$.
%This corresponds to the coupling between left and right-handed neutrinos
%being more suppressed
%for larger $R$, as the right-handed neutrino spends less
%of its time on the 3 space dimensional brane where the left-handed
%neutrino lives the larger $R$ becomes.
The Dirac neutrino mass is therefore suppressed relative to the
electron mass, and may be estimated as:
%\begin{equation}
$m_{LR}^{\nu}
%\sim \frac{m_e}{\sqrt{M_{string}R}}
\sim \frac{M_{string}}{M_{Planck}} m_e$
%\end{equation}
%where we have used Eq.\ref{flat} with $n=1$.
where $M_{Planck}\sim 10^{19}\ GeV/c^2$, and $M_{string}$ is the
string scale. Clearly low string scales, below the Planck scale,
can lead to suppressed Dirac neutrino masses. Similar suppressions
can be achieved with anisotropic compactifications
\cite{Antusch:2005kf}.

For the case of ``warped'' extra dimensions things are more
complicated/interesting \cite{Huber:2002sp}. Typically there are
two branes, a ``Planck brane'' and a ``TeV brane'', with all the
fermions and the Higgs in the ``bulk'' and having different
``wavefunctions'' which are more or less strongly peaked on the
TeV brane. The strength of the Yukawa coupling to the Higgs is
determined by the overlap of a particular fermion wavefunction
with the Higgs wavefunction, leading to exponentially suppressed
Dirac masses. For example the Higgs and top quark wavefunctions
are both strongly peaked on the TeV brane, leading to a large top
quark mass, while the neutrino wavefunctions will be strongly
peaked on the Planck brane leading to exponentialy suppressed
Dirac masses.

\section{What if Neutrinos are Majorana?}
We have already remarked that neutrinos, being electrically
neutral, allow the possibility of Majorana neutrino masses.
However such masses are forbidden in the SM since neutrinos form
part of a lepton doublet $L$, and the Higgs field also forms a
doublet $H$, and $SU(2)_L\times U(1)_Y$ gauge invariance forbids a
Yukawa interaction like $HLL$. So, if we want to obtain Majorana
masses, we must go beyond the SM.

One possibility is to introduce Higgs triplets $\Delta$ such that
a Yukawa interaction like $\Delta LL$ is allowed. However the
limit from the SM $\rho$ parameter implies that the Higgs triplet
should have a VEV $<\Delta > < 8$ GeV. One big advantage is that
the Higgs triplets may be discovered at the LHC and so this
mechanism of neutrino mass generation is directly testable
\cite{Perez:2008ha}.

Another possibility, originally suggested by Weinberg, is that
neutrino Majorana masses originate from operators $HHLL$ involving
two Higgs doublets and two lepton doublets, which, being higher
order, must be suppressed by some large mass scale(s) $M$. When
the Higgs doublets get their VEVs Majorana neutrino masses result:
%$\frac{\lambda_{\nu}}{M}<H>^2LL \rightarrow  m_{LL}^{\nu}\overline{\nu_L}\nu_L^c$
$m_{LL}^{\nu}=\lambda_{\nu}<H>^2/M$. This is nice because the
large Higgs VEV $<H>\approx  175$ GeV can lead to small neutrino
masses providing that the mass scale $M$ is high enough. E.g. if
$M$ is equal to the GUT scale $1.75.10^{16}$ GeV then
$m_{LL}^{\nu}=\lambda_{\nu}1.75.10^{-3}$ eV. To obtain larger
neutrino masses we need to reduce $M$ below the GUT scale (since
we cannot make $\lambda_{\nu}$ too large otherwise it becomes
non-perturbative).

Typically in physics whenever we see a large mass scale $M$
associated with a non-renormalizable operator we tend to associate
it with tree level exchange of some heavy particle or particles of
mass $M$ in order to make the high energy theory renormalizable
once again. This idea leads directly to the see-saw mechanism
where the exchanged particles can either couple to $HL$, in which
case they must be either fermionic singlets (right-handed
neutrinos) or fermionic triplets, or they can couple to $LL$ and
$HH$, in which case they must be scalar triplets. These three
possibilities have been called the type I, III and II see-saw
mechanisms, respectively.
%The type I and type II see-saw mechanisms are illustrated in Figs.\ref{I},\ref{II}.
%\begin{figure}[h]
%\begin{minipage}{10pc}
%\includegraphics[width=10pc]{typeI.eps}
%\caption{\label{I}Type I see-saw mechanism.}
%\end{minipage}\hspace{6pc}%
%\begin{minipage}{10pc}
%\includegraphics[width=10pc]{typeII.eps}
%\caption{\label{II}Type II see-saw mechanism.}
%\end{minipage}
%\end{figure}
If the coupling $\lambda_{\nu}$ is very small (for some reason)
then $M$ could even be lowered to the TeV scale and the see-saw
scale could be probed at the LHC \cite{King:2004cx}, however the
see-saw mechanism then no longer solves the problem of the
smallness of neutrino masses.

There are other ways to generate Majorana neutrino masses which
lie outside of the above discussion. One possibility is to
introduce additional Higgs singlets and triplets in such a way as
to allow neutrino Majorana masses to be generated at either one
\cite{Zee:1980ai} or two \cite{Babu:1988ki} loops. Another
possibility is within the framework of R-parity violating
Supersymmetry in which the sneutrinos $\tilde{\nu}$ get small VEVs
inducing a mixing between neutrinos and neutralinos $\chi$
%via diagrams
%similar to Fig.\ref{I}, but with sneutrinos replacing the Higgs and neutralinos replacing the
%right-handed neutrinos,
leading to Majorana neutrino masses $m_{LL}\approx
<\tilde{\nu}>^2/M_{\chi}$, where for example $<\tilde{\nu}>\approx
$ MeV, $M_{\chi}\approx $ TeV leads to $m_{LL}\approx $ eV. A
viable spectrum of neutrino masses and mixings can be achieved at
the one loop level \cite{Hirsch:2000ef}.

\section{Normal or Inverted?}
If the mass ordering is inverted then this may indicate a new
symmetry such as $L_e - L_{\mu} - L_{\tau}$ \cite{Petcov:1982ya}
or a $U(1)$ family symmetry \cite{King:2000ce}. However let us
assume that the hierarchy is normal and proceed down the road map
to the next experimental question.

\section{Very precise tri-bimaximal mixing?}
It is a striking fact that current data on lepton mixing is
(approximately) consistent with the so-called tri-bimaximal (TB)
mixing pattern \cite{Harrison:2002er},
\begin{equation}
\label{TBM}
U_{TB}= \left(\begin{array}{ccc} \sqrt{\frac{2}{3}}& \frac{1}{\sqrt{3}}&0\\
-\frac{1}{\sqrt{6}}&\frac{1}{\sqrt{3}}&\frac{1}{\sqrt{2}}\\
\frac{1}{\sqrt{6}}&-\frac{1}{\sqrt{3}}&\frac{1}{\sqrt{2}}
\end{array} \right) P_{Maj},
\end{equation}
where $P_{Maj}$ is the diagonal phase matrix involving the two
observable Majorana phases. However there is no convincing reason
to expect exact TB mixing, and in general we expect deviations.
These deviations can be parametrized by three parameters $r,s,a$
defined as \cite{King:2007pr}:
\begin{equation}
\sin \theta_{13} = \frac{r}{\sqrt{2}}, \ \ \sin \theta_{12} =
\frac{1}{\sqrt{3}}(1+s), \ \ \sin \theta_{23} =
\frac{1}{\sqrt{2}}(1+a). \label{rsa}
\end{equation}
Global fits of the conventional mixing angles
\cite{Schwetz:2008er,Fogli:2008jx} can be translated into the
$1\sigma$ ranges
\begin{equation}
0.14<r<0.24,\ \ -0.05<s<0.02, \ \ -0.04<a<0.10.
\end{equation}
Note in particular that the central value of $r$ is now 0.2 which
corresponds to a 2$\sigma$ indication for a non-zero reactor angle
as discussed at this meeting by Fogli \cite{Fogli:2008jx}.

Clearly a non-zero value of $r$, if confirmed, would rule out TB
mixing. However it is possible to preserve the good predictions
that $s=a=0$, by postulating a modified form of mixing matrix
called tri-bimaximal-reactor (TBR) mixing \cite{King:2009qt},
\begin{eqnarray}
U_{TBR} = \left( \begin{array}{ccc}
\sqrt{\frac{2}{3}}  & \frac{1}{\sqrt{3}} & \frac{1}{\sqrt{2}}re^{-i\delta } \\
-\frac{1}{\sqrt{6}}(1+ re^{i\delta })  & \frac{1}{\sqrt{3}}(1-
\frac{1}{2}re^{i\delta })
& \frac{1}{\sqrt{2}} \\
\frac{1}{\sqrt{6}}(1- re^{i\delta })  & -\frac{1}{\sqrt{3}}(1+
\frac{1}{2}re^{i\delta })
 & \frac{1}{\sqrt{2}}
\end{array}
\right)P_{Maj}. \label{MNS3}
\end{eqnarray}

Note that TBR mixing is distinct from the tri-maximal proposal
\cite{Wolfenstein:1978uw} that the second column of the mixing matrix should
consist of a column with all elements equal to $1/\sqrt{3}$.
On the one hand the TBR mixing proposal, to leading order in $r$, predicts the deviation parameters $s=a=0$ for
all $r$, whereas on the other hand the tri-maximal mixing proposal, to the same approximation, predicts
$s=0$ but $a=-(r/2)\cos \delta$. Thus, tri-bimaximal-reactor mixing and tri-maximal mixing
may be distinguished by accurate determinations of $a,r,\delta$
(i.e. $\theta_{23}$, $\theta_{13}$ and $\cos \delta$) at future
high precision neutrino facilities.

\section{Family Symmetry?}
Assuming that TB or TBR mixing is very precise and is not an
accident, it could be interpreted as a signal of an underlying
family symmetry. Indeed I am unaware of any viable alternative at
present. To understand the emergence of a family symmetry, let us
expand the neutrino mass matrix in the diagonal charged lepton
basis, assuming exact TB mixing, as $m_{LL}^{\nu}=U_{TB}{\rm
diag}(m_1, m_2, m_3)U_{TB}^T$ leading to (absorbing the Majorana
phases in $m_i$):
\begin{equation}
\label{mLL} m_{LL}^{\nu}=\frac{m_3}{2}\Phi_3 \Phi_3^T +
\frac{m_2}{3}\Phi_2 \Phi_2^T + \frac{m_1}{6}\Phi_1 \Phi_1^T
\end{equation}
where $\Phi_3^T=(0,1,1)$, $\Phi_2^T=(1,1,-1)$, $\Phi_1^T=(2,-1,1)$
and $m_i$ are the physical neutrino masses. This shows that the
neutrino mass matrix corresponding to TB mixing may be constructed
from the very simple orthogonal column vectors $\Phi_i$, whose
simplicity motivates an underlying non-Abelian family symmetry
involving all three families. The idea is that $\Phi_i$ are
promoted to new Higgs fields called ``flavons'' whose VEVs break
the family symmetry, with the particular vacuum alignments as
above. Such vacuum alignments can more readily be achieved if the
non-Abelian family symmetry is a discrete symmetry containing a
permutation symmetry capable of leading to $<|\Phi_2^T|>\propto
(1,1,1)$ \cite{Altarelli:2005yp}. A minimal choice of such family
symmetry seems to be $A_4$ \cite{Ma:2001dn} which only involves
the flavon $<|\Phi_2^T|>\propto (1,1,1)$ together with a further
flavon $<|\Phi_0^T|>\propto (0,0,1)$. Such minimal $A_4$ models
lead to neutrino mass sum rules between the three masses $m_i$,
resulting in/from a simplified mass matrix in Eq.\ref{mLL}. $A_4$
may result from 6D orbifold models \cite{Altarelli:2006kg}.

It is possible to derive the TB form of the neutrino mass matrix
in Eq.\ref{mLL} from the see-saw mechanism in a very elegant way
as follows. In the diagonal right-handed neutrino mass basis we
may write $M_{RR}^{\nu}={\rm diag}(M_A, M_B, M_C)$ and the Dirac
mass matrix as $m_{LR}^{\nu}=(A,B,C)$ where $A,B,C$ are three
column vectors. Then the type I see-saw formula
$m_{LL}^{\nu}=m_{LR}^{\nu}(M_{RR}^{\nu})^{-1}(m_{LR}^{\nu})^T$
gives
\begin{equation}
\label{mLLCSD} m_{LL}^{\nu}=\frac{AA^T}{M_A}+ \frac{BB^T}{M_B} +
\frac{CC^T}{M_C}.
\end{equation}
By comparing Eq.\ref{mLLCSD} to the TB form in Eq.\ref{mLL} it is
clear that TB mixing will be achieved if $A\propto \Phi_3$,
$B\propto \Phi_2$, $C\propto \Phi_1$, with each of $m_{3,2,1}$
originating from a particular right-handed neutrino of mass
$M_{A,B,C}$, respectively \cite{King:2005bj}. This mechanism
allows a completely general neutrino mass spectrum and, since the
resulting $m_{LL}^{\nu}$ is form diagonalizable, it
is referred to as form dominance (FD) \cite{Chen:2009um}. For
example, it has recently been show that the $A_4$ see-saw models \cite{Altarelli:2005yp} satisfy
FD \cite{Chen:2009um}.

If $m_1\ll m_2 < m_3$ then the precise form of $C$ becomes
irrelevant, and in this case FD reduces to constrained sequential
dominance (CSD)\cite{King:2005bj}. The CSD mechanism has been
applied in this case to models based on the family symmetries
$SO(3)$ \cite{King:2005bj,King:2006me} and $SU(3)$
\cite{deMedeirosVarzielas:2005ax}, and their discrete subgroups
\cite{deMedeirosVarzielas:2005qg}.

It is possible to achieve TBR mixing, corresponding to $s=a=0$ but
$r\neq 0$, by a slight
modification to the CSD conditions,
\begin{equation}
\label{PCSD} B = \frac{b}{\sqrt{3}} \left(
\begin{array}{r}
1 \\
1 \\
-1
\end{array}
\right),\ \ A  = \frac{c}{\sqrt{2}} \left(
\begin{array}{r}
\varepsilon \\
1 \\
1
\end{array}
\right).
\end{equation}
We refer to this as Partially Constrained Sequential Dominance
(PCSD)\cite{King:2009qt}, since one of the conditions of CSD is
maintained, while the other one is violated by the parameter
$\varepsilon$. Note that the introduction of the parameter
$\varepsilon$ also implies a violation of FD since the columns of
the Dirac mass matrix $A,B$ can no longer be identified with the
columns of the MNS matrix, due to the non-orthogonality of $A$ and
$B$. To leading order in $|m_2|/|m_3|$ the mass matrix resulting
from Eq.\ref{PCSD} leads to TBR mixing where we identify \cite{King:2009qt},
\begin{equation}
m_1=0, \ \ m_2=b^2/M_B, \ \ m_3=a^2/M_A, \ \ \varepsilon =
re^{-i\delta }.
\end{equation}
Thus, the TBR form of mixing matrix in
Eq.\ref{MNS3} will result, to leading order in $|m_2|/|m_3|$.

\section{Hierarchical or Degenerate?}
This key experimental question may be decided by the same
experiments as will also determine the nature of neutrino mass
(Dirac or Majorana) \cite{4}. Although not a theorem, it seems
that a hierarchical spectrum could indicate a type I see-saw
mechanism, while a (quasi) degenerate spectrum could imply a type
II see-saw mechanism. It is possible that a type II see-saw
mechanism could naturally explain the degenerate mass scale with
the degeneracy enforced by an $SO(3)$ family symmetry, while the
type I see-saw part could be responsible for the small neutrino
mass splittings and the (TB) mixing \cite{Antusch:2004xd}.
An $A_4$ model of quasi-degenerate neutrinos with TB mixing, working at the
effective neutrino mass operator level, was considered recently in
\cite{Varzielas:2008jm}.

\section{GUTs and/or Strings?}
Finally we have reached the end of the decision tree, with the
possibility of an all-encompassing unified theory of flavour based
on GUTs and/or strings. Such theories could also include a family
symmetry in order to account for the TB mixing. There are many
possibilities for the choice of family symmetry and GUT symmetry.
Examples include the Pati-Salam gauge group $SU(4)_{PS}\times
SU(2)_L\times SU(2)_R$ in combination with $SU(3)$
\cite{deMedeirosVarzielas:2005ax}, $SO(3)$
\cite{King:2005bj,King:2006me}, $A_4$ \cite{King:2006np} or
$\Delta_{27}$ \cite{deMedeirosVarzielas:2006fc}. Other examples
are based on $SU(5)$ GUTs in combination with $A_4$
\cite{Altarelli:2008bg} or $T'$ \cite{Chen:2007afa}.

For example, it is straightforward to implement the above example of PCSD into realistic GUT models
with non-Abelian
family symmetry spontaneously broken by flavons which are based on the CSD
mechanism \cite{King:2005bj,deMedeirosVarzielas:2005ax,King:2006me}.
In such models the columns of the Dirac mass matrix in the diagonal
right-handed neutrino mass basis are determined by flavon vacuum alignment,
with the column $B$ identified with a triplet flavon $\phi_{123}$ and
the column $A$ identified with a triplet flavon $\phi_{23}$
and it is quite easy to obtain a correction to the vacuum aligmment
such that
\begin{equation}
\label{PCSD4}
\langle \phi_{123} \rangle  = \frac{b}{\sqrt{3}}
\left(
\begin{array}{r}
1 \\
1 \\
-1
\end{array}
\right),\ \
\langle \phi_{23} \rangle  = \frac{a}{\sqrt{2}}
\left(
\begin{array}{r}
\varepsilon \\
1 \\
1
\end{array}
\right),
\end{equation}
in direct correspondence with Eq.\ref{PCSD}.
For example, in such models based on the discrete family symmetry $A_4$
\cite{King:2006me},
the flavon vacuum expectation
value (VEV) $\langle \phi_{123} \rangle $ will preserve a $Z_2$
subgroup of the original discrete family symmetry corresponding to an $A_4$
generator $S$ \cite{Altarelli:2006kg}, while the flavon VEV $\langle \phi_{23} \rangle $ will
violate this subgroup even in the limit that $\varepsilon=0$. It is therefore
natural to assume some misalignment of $\langle \phi_{23} \rangle $ since,
unlike $\langle \phi_{123} \rangle $, it is not protected by any symmetry.

In typical Family Symmetry $\otimes$ GUT models the origin of the
quark mixing angles derives predominantly from the down quark
sector, which in turn is closely related to the charged lepton
sector. In order to reconcile the down quark and charged lepton
masses, simple ansatze, such as the Georgi-Jarlskog hypothesis
\cite{Georgi:1979df}, lead to very simple approximate expectations
for the charged lepton mixing angles such as $\theta^e_{12}\approx
\lambda/3$, $\theta^e_{23}\approx \lambda^2$,
$\theta^e_{13}\approx \lambda^3$, where $\lambda \approx 0.22$ is
the Wolfenstein parameter from the quark mixing matrix. If the
family symmetry enforces accurate TB mixing in the neutrino
sector, then $\theta^e_{12}\approx \lambda/3$ charged lepton
corrections will cause deviations from TB mixing in the physical
lepton mixing angles, and lead to a sum rule relation
\cite{King:2005bj,Masina:2005hf,Antusch:2005kw}, which can be
conveniently expressed as \cite{King:2007pr} $s\approx r \cos
\delta$ where $r\approx \lambda /3$ and $\delta$ is the observable
CP violating oscillation phase, with RG corrections of less than
one degree \cite{Boudjemaa:2008jf}. Such sum rules can be tested
in future high precision neutrino oscillation experiments
\cite{Antusch:2007rk}.

Note that in such a GUT-flavour framework, one expects the
charged lepton corrections to the neutrino mixing angles to be less than of order
$\theta_{12}^e/\sqrt{2}$
(where typically $\theta_{12}^e$ is a third of the Cabibbo angle)
plus perhaps a further $1^o$ from renormalization group (RG) corrections.
Thus such theoretical corrections cannot account for an observed reactor angle
as large as $8^o$, corresponding to $r=0.2$,
starting from the hypothesis of exact TB neutrino mixing.

%\begin{figure}[h]
%\begin{minipage}{14pc}
%\includegraphics[width=14pc]{family_groups.eps}
%\caption{\label{family}Some possible family symmetry groups.}
%\end{minipage}\hspace{2pc}%
%\begin{minipage}{14pc}
%\includegraphics[width=14pc]{gut_groups.eps}
%\caption{\label{GUT}Some possible GUT groups.}
%\end{minipage}
%\end{figure}

\section{Conclusion}
Neutrino mass and mixing clearly requires new physics beyond the
SM, but in which direction should we go? There are many roads for
model building, but we have seen that answers to key experimental
questions will provide the sign posts {\it en route} to a unified
theory of flavour.

In particular we would like to emphasize that a measurement of a
large reactor angle, consistent with the present
$2\sigma$ indication for $r=0.2$, can still be consistent with tri-bimaximal solar and atmospheric
mixing, corresponding to $s=a=0$,
according to the tri-bimaximal-reactor mixing hypothesis.
By contrast, tri-maximal mixing predicts
$s=0$ but $a=-(r/2)\cos \delta$.

Alternatively the presence of a large reactor angle could be a sign that
TB mixing is the wrong starting point and instead it is better to
start from bi-maximal neutrino mixing and then
invoke large charged lepton corrections to correct the reactor and solar angles,
as discussed by Altarelli at this meeting \cite{Altarelli:2009gn}.

The common feature of all these approaches is the presence of an underlying family
symmetry, even though the reactor angle may be quite large. The existence of
such disparate approaches only underlines the need for further high precision
data from the neutrino experiments in order to resolve which approach is correct.

\section{Acknowledgements}
I would like to thank the organizers of the
XIII International Workshop on
"Neutrino Telescopes", Venice, March 10-13, 2009, Venice
for their kind hospitality, and for partial
support from the following grants: STFC Rolling Grant
ST/G000557/1; EU Network MRTN-CT-2004-503369; EU ILIAS RII3-CT-2004-506222.

\end{document}